\documentclass[reprint,aps,pre,showpacs,superscriptaddress,amsmath,onecolumn,11pt]{revtex4-1}
\usepackage{amssymb}
\usepackage{amsmath}
\usepackage{graphicx}

\usepackage[usenames,dvipsnames]{color}
\usepackage[normalem]{ulem}

\newcommand{\reffig}[1]{Fig.~\ref{#1}}
\begin{document}


\title{Self-organised dip-coating patterns of simple, partially wetting, nonvolatile liquids} 


\author{Walter Tewes}
\email[]{walter.tewes@uni-muenster.de}
\affiliation{Institut f\"ur Theoretische Physik, Westf\"alische Wilhelms-Universit\"at M\"unster, Wilhelm-Klemm-Str.\ 9, 48149 M\"unster, Germany}
\author{Markus Wilczek}
\author{Svetlana V. Gurevich}
\author{Uwe Thiele}
\email{u.thiele@uni-muenster.de}
\homepage{http://www.uwethiele.de}
\affiliation{Institut f\"ur Theoretische Physik, Westf\"alische Wilhelms-Universit\"at M\"unster, Wilhelm-Klemm-Str.\ 9, 48149 M\"unster, Germany}
\affiliation{Center for Nonlinear Science (CeNoS), Westf{\"a}lische Wilhelms-Universit\"at M\"unster, Corrensstr.\ 2, 48149 M\"unster, Germany}


\date{\today}

\begin{abstract}
When a solid substrate is withdrawn from a bath of simple, partially wetting, nonvolatile liquid, one typically distinguishes two regimes, namely, after withdrawal the substrate is macroscopically dry or homogeneously coated by a liquid film. In the latter case, the coating is called a Landau--Levich film. Its thickness depends on the angle and velocity of substrate withdrawal. We predict by means of a numerical and analytical investigation of a hydrodynamic thin-film model the existence of a third regime. It consists of the deposition of a regular pattern of liquid ridges oriented parallel to the meniscus. We establish that the mechanism of the underlying meniscus instability originates from competing film dewetting and Landau--Levich film deposition. Our analysis combines a marginal stability analysis, numerical time simulations and a numerical bifurcation study via path-continuation.
\end{abstract}


\maketitle

\section{Introduction}
\begin{figure}[h!]
  \centering
 \includegraphics[width=0.5\textwidth]{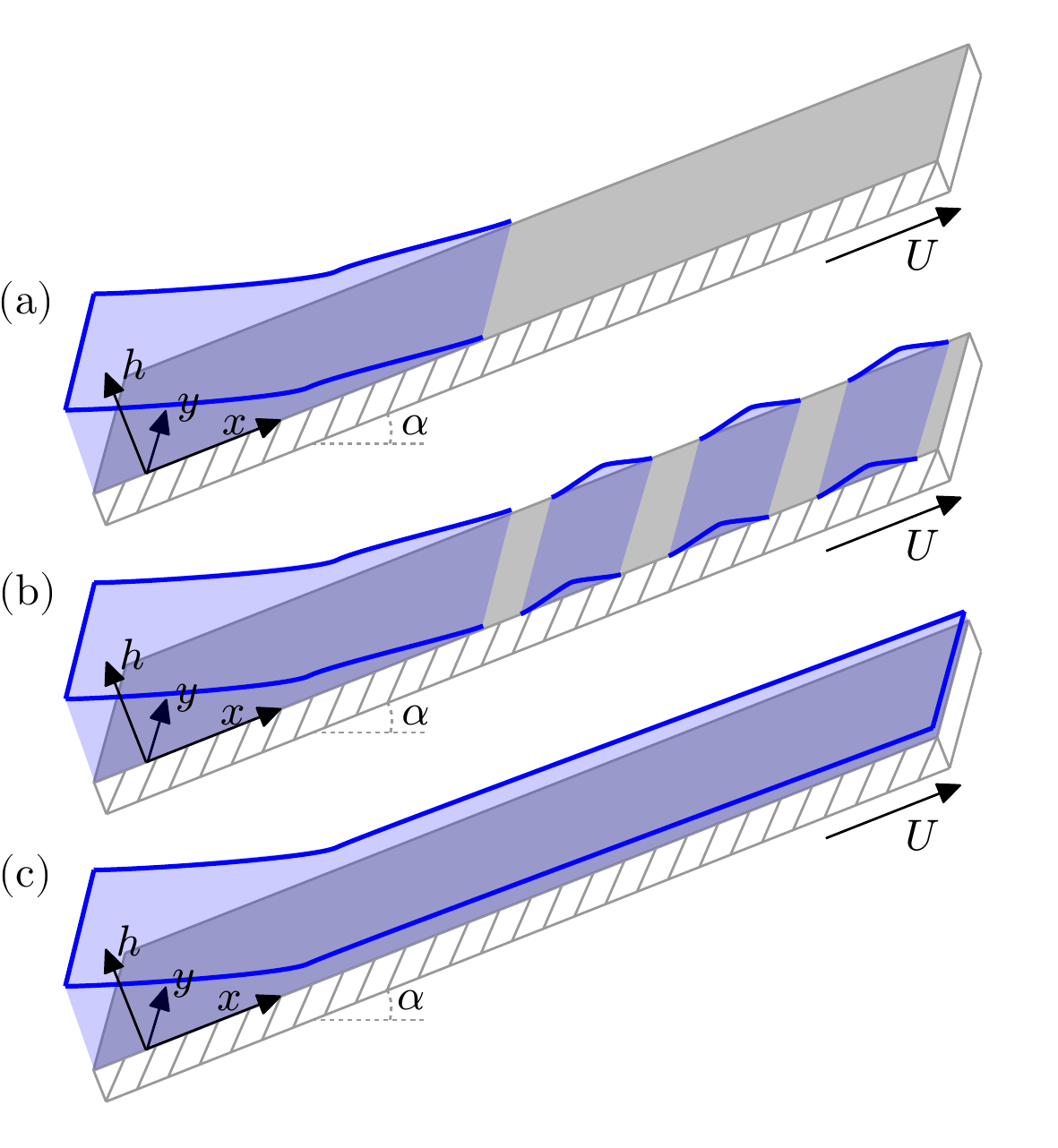}
 \caption{(Color online) Sketch of the investigated geometry: A solid substrate is withdrawn under an angle $\alpha$ and at a velocity $U$ from a bath of simple, partially wetting, nonvolatile liquid. Three main regimes are observed: (a) At small transfer velocities $U$, a liquid meniscus of finite length is formed and a macroscopically dry substrate emerges, while (c) at high velocities a Landau--Levich film is transferred. In the transition region, (b), at intermediate $U$, a stripe-like periodic deposition pattern can be observed.}
 \label{fig:fig1_sketch}
 \end{figure}

When a plate is withdrawn from a bath of simple, partially wetting, nonvolatile liquid, two distinct phenomena occur depending on the velocity of withdrawal $U$ \cite{SnAn2013arfm}: At low velocities, no macroscopic liquid film is transferred onto the moving plate and a static meniscus forms at the bath surface \cite{ChSE2012pf} (see sketch Fig.~\ref{fig:fig1_sketch}(a)). However, as studied in the seminal works by Landau and Levich \cite{LaLe1942apu} and by Derjaguin \cite{Derj1945apu}, at higher velocities a liquid film of well-defined height is transferred onto the plate. This occurs above the so-called \textit{Landau--Levich transition} (cf. Fig.~\ref{fig:fig1_sketch}(c)) and is the usual situation employed in coating applications \cite{Wils1982jem,KhKS1992ces,WeRu2004arfm}. 
The film height $h_\mathrm{f}$ of this so-called \textit{Landau--Levich film} is determined by the velocity of withdrawal $U$, the viscosity of the liquid, and its surface tension. Recently, the behavior of simple, partially wetting, nonvolatile liquids on inclined plates was experimentally \cite{SZAF2008prl,DFSA2008jfm} and theoretically \cite{SDFA2006prl,SADF2007jfm,ZiSE2009epjt,GTLT2014prl,TsGT2014epje,GLFD2016jfm} studied for velocities of withdrawal in the vicinity of the Landau--Levich transition.  
Further studies of Landau--Levich films for wetting liquids are found in \cite{Wils1982jem,JiAM2005pf,BeZu2008jfm,SZAF2008prl,BCMO2010jfm,MRRQ2011jcis}. Landau--Levich films and the related transition are also relevant for other geometries and driving forces, e.g., they occur at menisci driven by surface-acoustic waves \cite{MoMa2017jfm}, in rimming and coating flows, respectively, in and on rotating cylinders \cite{Wils1982jem,Melo1993pre,WiWi1997pf,SeTh2011rmp} and for the Bretherton problem of a gas bubble moving through a liquid-filled tube \cite{Bret1961jfm,TeDS1988rpap}.

In the present work, we focus on a scenario which has not yet been described, namely the self-organised deposition of a periodic array of liquid ridges from a meniscus onto the moving plate (see Fig.~\ref{fig:fig1_sketch}(b)). We show that this type of behavior can be found for liquids which exhibit spinodal dewetting \cite{Mitl1993jcis,BeNe2001prl} at film heights in the range of the thicknesses $h_\mathrm{f}$ of the dynamically produced Landau--Levich films. To investigate the effect, we employ a well-studied mesoscopic hydrodynamic thin-film model for the withdrawal of a plate from a simple, partially wetting, nonvolatile liquid \cite{GTLT2014prl}. At equilibrium, it allows for the coexistence of a macroscopic drop and a microscopic adsorption layer (precursor film) and is therefore well suited to describe the dynamics behaviour in the vicinity of the Landau--Levich transition as there the deposited coating changes from a microscopic adsorption layer that is macroscopically ``dry'' to a Landau--Levich coating film. The implied continuous character of the film surface profile across the contact-line region allows for detailed bifurcation studies and the identification of a variety of transition scenarios \cite{GTLT2014prl,TsGT2014epje,wilczek2016dip}.

Conceptually, such a deposition of periodic arrays of stripes or ridges belongs to a class of pattern formation processes triggered by a moving front as in the frame of the moving plate the meniscus can be considered as a front that moves with velocity $-U$. This externally imposed front velocity is a control parameter of the system. Examples for investigations of such triggered pattern formation range from the experimentally and theoretically investigated structure formation in Langmuir--Blodgett films \cite{KGFC2010l,KGFT2012njp,kopf2012substrate,KoTh2014n,WiGu2014pre,WTGK2015mmnpb}, over the study of Cahn--Hilliard-type model equations in one (1D) and two (2D) dimensions for externally quenched phase separation (e.g., by a moving temperature jump for films of polymer blends or binary mixtures) \cite{FoWa2009pre,FoWa2012pre,Krek2009pre} to the rigorous mathematical analysis of trigger fronts in a complex Ginzburg--Landau equation as well as in a Cahn--Hilliard and an Allen--Cahn equation \cite{GoSc2014jns,GoSc2015arma,monteiro2017phase}. In the aforementioned systems, a switch from a linearly stable to an unstable state takes place at a certain position within the considered domain. This position is moved with a constant velocity which is (directly or indirectly) controlled, leading to intriguing pattern formation phenomena.

The outline of the present work is as follows: First, in Section~\ref{sec:mm} we briefly present the thin-film model including the pertinent boundary conditions and outline the employed numerical methods. Next, in Section~\ref{SEC.LLTransition} we present the relevant bifurcation diagrams for the Landau--Levich transition and discuss the various branches of different types of linearly stable and unstable surface profiles that exist near the transition. 
Then, Section~\ref{sec:periodic} presents numerical time simulations and shows that there exists a regime where periodic structures form at the meniscus, i.e., a periodic array of liquid ridges parallel to the meniscus is deposited onto the moving plate. A phase diagram is obtained for these states that shows their existence in an extended region in parameter space. Section~\ref{SEC.MarginalStabilityAnalysis} employs a marginal stability analysis to determine the upper  velocity threshold for the formation of periodic structures as the linear propagation velocity of a dewetting front into spinodally unstable Landau--Levich films. The subsequent Section~\ref{SEC:BifStuctureOfPeriodic} employs an advanced path continuation technique to determine the full bifurcation structure including steady and time-periodic states. These are related to the results of the marginal stability analysis and numerical time simulations for 2D substrates. For the latter, Section~\ref{SEC.2D} compares the structures formed by meniscus-guided dewetting to dewetting structures found on horizontal substrates. Section~\ref{sec:conc} concludes with an outlook.

\section{Mathematical Model}
\label{sec:mm}

We model the evolution of the local height $h(\mathbf{r},t)$, $\mathbf{r}=(x,y)$ of the film of simple, nonvolatile  liquid in the laboratory frame for the dip-coating geometry shown in \reffig{fig:fig1_sketch} with the well-established long-wave, thin-film or lubrication equation \cite{OrDB1997rmp,CrMa2009rmp} in its nondimensional form, including a Derjaguin (or disjoining) pressure accounting for partial wettability \cite{StVe2009jpm,Thie2010jpcm}, gravitational contributions and an advection term accounting for the withdrawal of the substrate \cite{GTLT2014prl,wilczek2016dip}:

\begin{align}
 \partial_t h(\mathbf{r},t)&=\nabla\cdot\left[Q(h)\nabla\Big(-\Delta h+f_{\mathrm{D}}'(h)\Big)+\boldsymbol{\chi}(h)\right]-\mathbf{u}\cdot\nabla h, \label{EQ.tfincline}\\
 \mathrm{where~~}\boldsymbol{\chi}(h)&=G\,Q(h)\Big(\nabla h+\boldsymbol{\alpha}\Big),~~ \mathbf{u}=(U,0)^{\mathrm{T}},~~,\boldsymbol{\alpha}=(\alpha,0)^{\mathrm{T}},~~Q(h)=\frac{h^3}{3}.\label{EQ.tfincline1}
\end{align}
Here, $\alpha$ is the inclination angle of the substrate in long-wave scaling, $G$ a dimensionless gravity parameter and $Q(h)$ is the mobility resulting for a no-slip boundary condition at the substrate \cite{OrDB1997rmp}. For the specific scaling used in \eqref{EQ.tfincline} and \eqref{EQ.tfincline1} see \cite{GTLT2014prl}. We use the specific wetting potential \cite{Pism2001pre, Thie2010jpcm}
\begin{align}
f_{\mathrm{D}}(h)=-\frac{1}{2h^2}+\frac{1}{5h^5}
\label{EQ.disjoining}
\end{align}
resulting in the Derjaguin pressure $-f_{\mathrm{D}}'(h)=-1/h^3+1/h^6$.

Equation~(\ref{EQ.tfincline}) is in general defined on a two-dimensional domain $\left[0,L_x\right]\times\left[0,L_y\right]$, where we choose the first dimension to correspond to the direction of withdrawal (cf.~\reffig{fig:fig1_sketch}).
The boundary conditions, which are also employed in \cite{wilczek2016dip}, read
\begin{align}
 h=h_l, \nabla h=(-\alpha,0)^T,~~ \mathbf{n}\cdot\nabla \Delta h=0 ~~~~~&\mathrm{at}~~~x=0\label{EQ.bcdownstream}\\
 \mathbf{n}\cdot\nabla h=0,~~ \mathbf{n}\cdot\nabla \Delta h=0 ~~~~~&\mathrm{at}~~~x=L_x,~~\mathbf{n}=(1,0)^{\mathrm{T}}\label{EQ.bcupstream}\\
 h(x,y=0)=h(x,y=L_y).
\end{align}
The downstream boundary conditions \eqref{EQ.bcdownstream} at $x = 0$ model the smooth transition to the liquid bath, while the upstream boundary condition at $x=L_x$ should permit the liquid to flow out of the considered domain.
The value of $h_l$ in \eqref{EQ.bcdownstream} is chosen large enough that one can safely neglect the next terms of the asymptotic series derived via central manifold projection in \cite{TsGT2014epje}.

\begin{figure}[h!]
  \centering
 \includegraphics[width=0.75\textwidth]{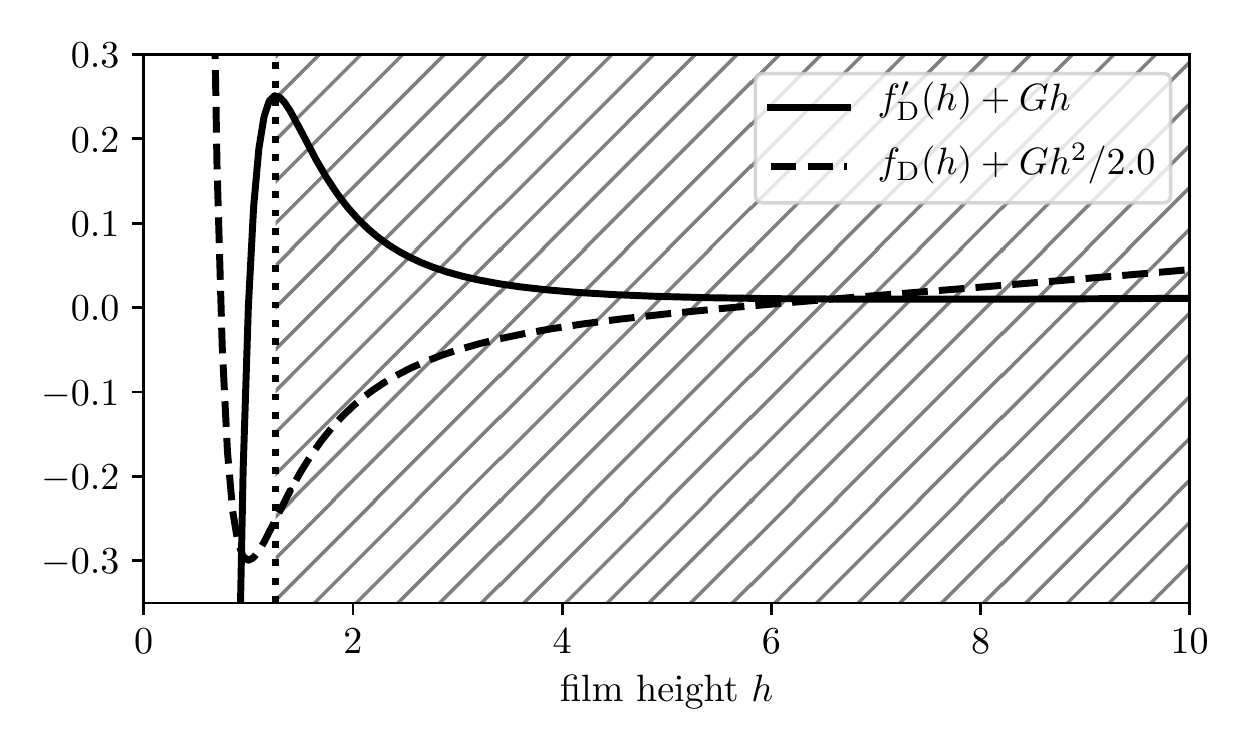}
 \caption{Effective local wetting energy $f(h)$ and derived pressure for the model wetting potential \eqref{EQ.disjoining} at $G=0.001$. The dotted line indicates the threshold film height, above which flat films are linearly unstable until they are eventually stabilized by gravity.}
 \label{fig:fig2_disjoining}
\end{figure}

In the presence of gravity, the hydrostatic term in $\boldsymbol{\chi}(h)$ can be combined with the wetting potential to give the effective local energy (cf.~\reffig{fig:fig2_disjoining}):
\begin{align}
 f(h)=f_{\mathrm{D}}(h)+\frac{G}{2}h^2. \label{EQ.effectiveLocalEnergy}
\end{align}
It exhibits a minimum at a small film height $h_a$ and thus for a horizontal substrate a macroscopic drop can coexist with a stable thin adsorption layer of height $h_{a}$. The corresponding region outside the drop can be considered as macroscopically ``dry'' (see, e.g., the case of pure liquid in \cite{TSTJ2018l}). This corresponds to the ``moist'' case in \cite{deGennes1985rmp}. This implies for the present dragged meniscus setting that there such an adsorption layer always exists upstream of the meniscus, even below the Landau--Levich transition \cite{GTLT2014prl}. This is in contrast to the slip model in \cite{SADF2007jfm,ZiSE2009epjt}.
Here, the equilibrium contact angle $\theta_{\mathrm{eq}}$ in the long-wave scaling is related to the energy $f(h)$ through $\theta_\mathrm{eq}=\sqrt{2\left|f(h_a)\right|}$ and is for the employed $G\ll1$ practically identical to the case without gravity. 

Next, we discuss the linear stability of a flat film of thickness $h_0$ on a resting inclined homogeneous substrate. This is done by inserting the ansatz $h(x,t)=h_0+\varepsilon~\mathrm{e}^{ikx+\omega(k)t}$ into \eqref{EQ.tfincline} for $U=0$, expanding in $\varepsilon\ll1$. To order $\mathcal{O}(\varepsilon)$, we obtain the dispersion relation
\begin{align}
 \omega(k)=-Q(h_0)k^2\left[k^2+f''(h_0)\right]+ikG\alpha Q'(h_0). \label{EQ.dispersionRelation}
\end{align}
that determines the development of harmonic perturbation modes of small amplitude.
For $f''(h_0)<0$, the film is unstable and develops a surface instability that results in spinodal dewetting \cite{Mitl1993jcis,ThVN2001prl}. For the particular local energy \eqref{EQ.effectiveLocalEnergy}, liquid films are linearly stable below a critical film height slightly larger than the adsorption layer height $h_a$. Then, short-range stabilising contributions dominate over the long-range destabilising terms. Mesoscopic film heights above this threshold are unstable (cf.~\reffig{fig:fig2_disjoining}). Only at larger film heights $h \approx (3/G)^{1/4}$, flat films get eventually stabilised by the hydrostatic contribution. 

Our numerical investigations of the equations \eqref{EQ.tfincline}-\eqref{EQ.bcupstream} are conducted by two complementary approaches. Stable and unstable steady profiles as well as time-periodic profiles together with their bifurcation diagrams are calculated by numerical pseudo-arclength continuation \cite{DWCD2014ccp,EGUW2018preprint} using the package auto07p \cite{DoKK1991ijbc,DoKK1991ijbcb}. Formulating the 1D version of \eqref{EQ.tfincline} with $\partial_t h=0$ as a small system of three first order ordinary differential equations (ODE) in space (after one integration in space), the parameter continuation allows us to obtain bifurcation diagrams of the steady film profiles which are briefly discussed in Section \ref{SEC.LLTransition}. For a recent review of such  techniques cf. \cite{EGUW2018preprint}, tutorials can be found in \cite{cenosTutorial}. However, if we employ a spatial finite difference approximation of \eqref{EQ.tfincline}-\eqref{EQ.bcupstream}, we are able to formulate the full time-dependent problem as a rather large system of first order ODE in time. This allows us to also determine stable and unstable time-periodic states by continuation although for a reduced domain size (cf.~Section \ref{SEC:BifStuctureOfPeriodic}). For further implementation details see \cite{KoTh2014n,LRTT2016pf} where similar techniques are employed for thin-film equations and closely related driven Cahn--Hilliard equations.

To investigate the time-periodic states, we furthermore conduct numerical time simulations using a finite element method. The approach uses quadratic elements with linear (Q1) ansatz and test functions as well as an implicit second-order Runge--Kutta time stepping scheme. They are implemented using the open-source framework DUNE-PDELab \cite{bastian2008genericI,bastian2008genericII,bastian2010generic}. For further implementation details, see also \cite{wilczek2016dip}, where a similar numerical approach is employed. One-dimensional simulations are performed on a domain $\Omega_\mathrm{1D} = [0,400]$ discretised on 400 elements, while two dimensional simulations are performed on $\Omega_\mathrm{2D} = [0,600]\times[0,300]$ discretised on $300\times150$ elements.

\section{Steady States: The Landau--Levich Transition} \label{SEC.LLTransition}

\begin{figure}[h!]
  \centering
 \includegraphics[width=0.9\textwidth]{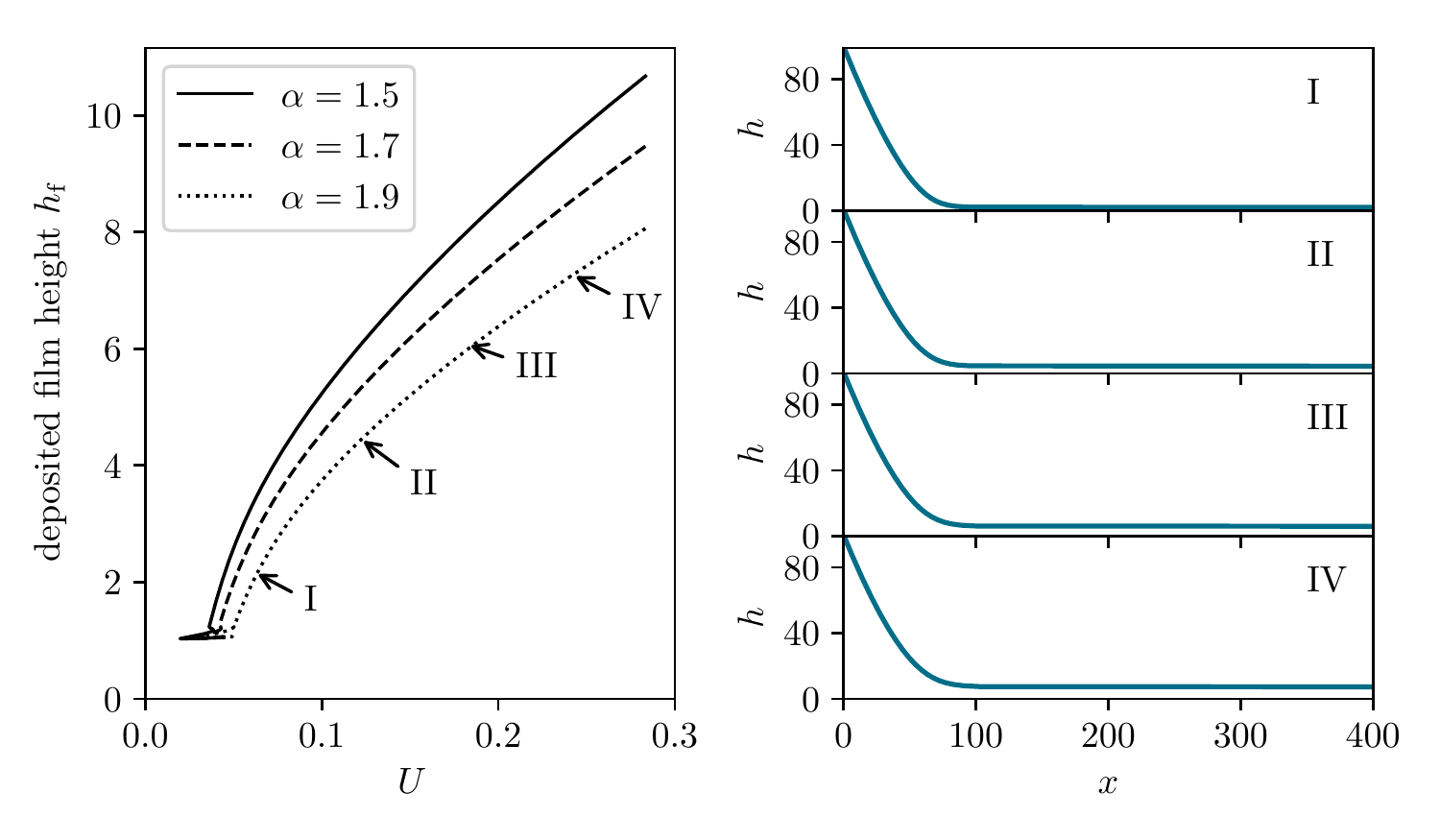}
 \caption{(left) Bifurcation diagram showing the height  of the transferred film $h_\mathrm{f}$ in dependence of the velocity of plate withdrawal $U$ for different inclination angles $\alpha$ as given in the legend. (right) Examples of Landau--Levich film height profiles for $\alpha=1.9$ at velocities indicated by corresponding roman numbers in the left panel.}
 \label{fig:fig3_bifurcation_LL}
\end{figure}

\begin{figure}[ht!]
  \centering
 \includegraphics[width=0.9\textwidth]{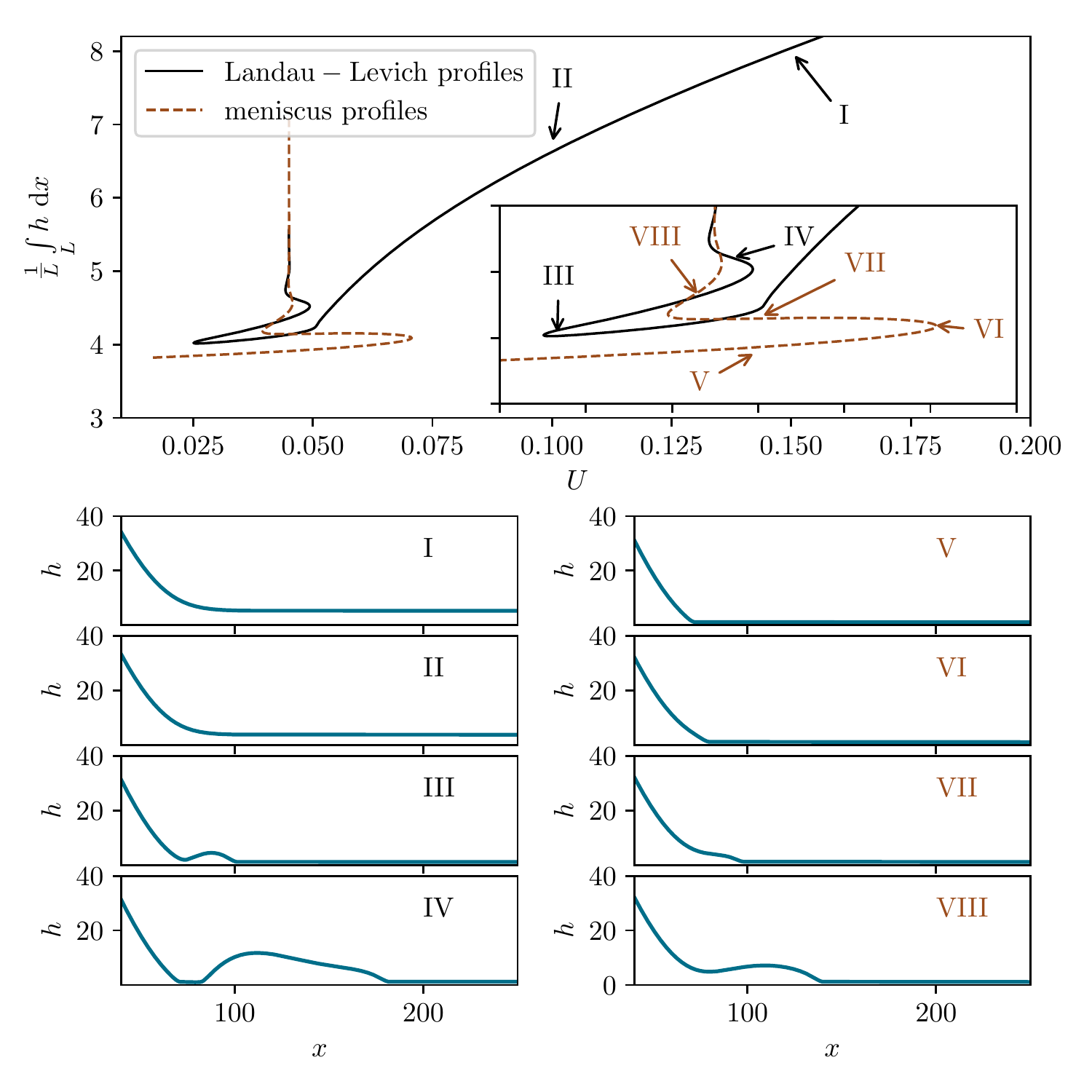}
 \caption{(top) Bifurcation diagram showing the average film height of steady state solutions of equations \eqref{EQ.tfincline}-\eqref{EQ.bcupstream} in dependence of velocity $U$ of plate withdrawal at fixed inclination $\alpha=2.0$. The red dashed, exponentially snaking branch corresponds to meniscus- and attached-foot solutions, the black solid branch corresponds at high velocities to Landau--Levich film states and becomes at lower $U$ a snaking curve of detached-foot solutions close to the Landau--Levich transition. (bottom) Examples of steady profiles at points on the two branches indicated by corresponding roman numbers in the top panel.}
 \label{fig:fig4_bifurcation_LL_foot}
\end{figure}

First, we discuss steady solutions of equations \eqref{EQ.tfincline}-\eqref{EQ.bcupstream} in 1D as obtained by numerical continuation. In \reffig{fig:fig3_bifurcation_LL}, the emergence of Landau--Levich film solutions at a critical velocity of plate withdrawal is illustrated. In particular, we show the height of the transferred film $h_\mathrm{f}$ in dependence of the velocity of plate withdrawal for three different inclinations $\alpha$. Above the Landau--Levich  transition that occurs at about $U=0.04$, $h_\mathrm{f}$ monotonically increases following the well-known scaling law derived in \cite{LaLe1942apu}:
\begin{align}
h_\mathrm{f}\propto U^{2/3}.
\end{align}
\reffig{fig:fig4_bifurcation_LL_foot} further illustrates that due to the interplay of wetting potential, plate inclination and dragging by the moving plate the details of the transition to the Landau--Levich film can be rather complex and strongly depend on the inclination angle (cf.~\cite{GTLT2014prl} for a discussion in the context of unbinding transitions). The lower panels of \reffig{fig:fig4_bifurcation_LL_foot} give a selection of different stable and unstable steady profile types occurring for $\alpha=2.0$ in the vicinity of the Landau--Levich  transition. The upper panel gives the corresponding bifurcation diagram where as
solution measure we use the mean amount of liquid per area. In contrast to the cases shown in \cite{GTLT2014prl}, the branch of Landau--Levich films (cf.\ profiles I and II) is not connected to the meniscus states but to a snaking branch of \textit{detached-foot solutions} where an extended foot is connected to the meniscus by a finite length of a thin nearly flat film (e.g., profiles III and IV in \reffig{fig:fig4_bifurcation_LL_foot}). 
In experimental studies \cite{SZAF2008prl}, similar profiles were found to detach from the meniscus as the speed $U$ is increased above the Landau--Levich threshold.
Furthermore, another snaking branch of standard {\it{attached-foot solutions}} exists without deep depression between foot and meniscus. Note, however that the film height does not need to be monotonic (see profiles V-VIII). This branch is not connected to the Landau--Levich film branch but to the branch of meniscus solutions. This branch and the corresponding attached-foot solutions are extensively investigated with slip and precursor models in Refs.~\cite{GTLT2014prl,TsGT2014epje,SZAF2008prl,ZiSE2009epjt}. For a detailed study of solution branches and their reconnections with variation of the inclination see \cite{Galvagno2015-phd}. Here, the study of the steady states sets the stage for our main focus that lies on time-periodic states. 
The latter have to our knowledge not yet been described in the literature. In particular, the steady Landau--Levich films discussed at \reffig{fig:fig3_bifurcation_LL} play a crucial role in the theoretical analysis of the periodic deposition structures and are furthermore used as initial conditions for the numerical time simulations.


\section{Periodic Depositions}
\label{sec:periodic}
\begin{figure}[tbh]
  \centering
 \includegraphics[width=0.9\textwidth]{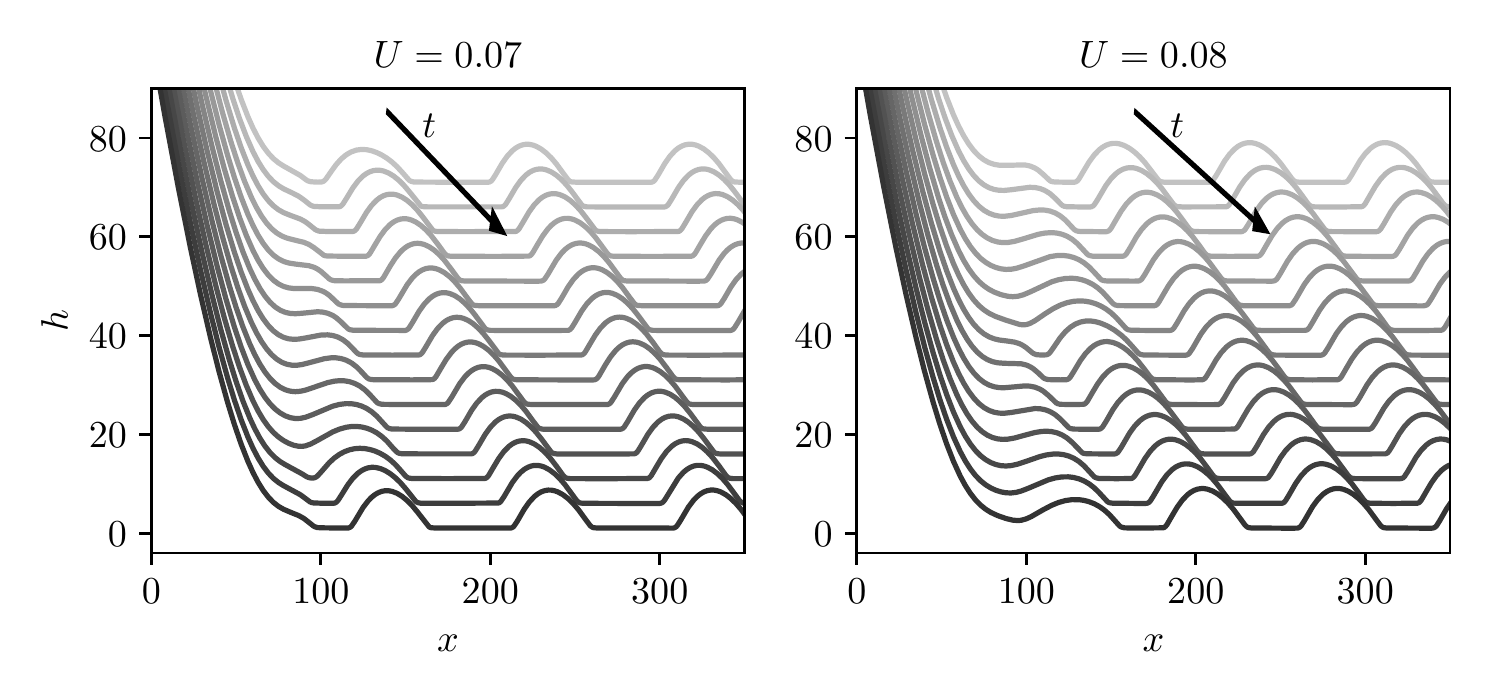}
 \caption{Sequences of snapshots of height profiles $h(x,t)$ arranged in the form of space-time plots. They result from numerical time simulations of \eqref{EQ.tfincline}-\eqref{EQ.bcupstream} at inclination $\alpha=2.0$ with (left) $U=0.07$ and (right) $U=0.08$. A time-periodic meniscus dynamics results in the periodic detachment of liquid ridges and, in consequence, in the transfer of a spatially periodic structure onto the moving plate. The time evolution is visualised by a vertical offset of $\Delta h=5$ between subsequent snapshots that lie $\Delta t=200$ apart.}
 \label{fig:fig5_snapshots_1D}
\end{figure}
 
Next, we perform time simulations of equations \eqref{EQ.tfincline}-\eqref{EQ.bcupstream} in 1D. As initial condition, we either use a steady meniscus profile or a steady Landau--Levich film obtained in Section \ref{SEC.LLTransition} by numerical continuation. At given fixed inclination angle $\alpha$, the Landau--Levich film is always stable at large velocities $U$. Similarly, for sufficiently small velocities $U$, a stable meniscus solution is observed. However, in a certain intermediate velocity range at about and not too far above the Landau--Levich transition, we find that the system converges towards stable time-periodic states. They correspond to a time-periodic detachment of liquid ridges from the meniscus, i.e., at constant frequency. Then, the ridges move as a spatially periodic stripe-like pattern up with the moving substrate (but not at the same velocity). 
\reffig{fig:fig5_snapshots_1D} shows space-time plots illustrating the process for $U=0.07$ and $U=0.08$. Note that the respective arrow faithfully indicates the motion of a point fixed to the moving substrate. This clearly indicates that the ridges travel slower than the plate once they are detached from the meniscus. This is because the ridges slowly slide down the plate due to gravity.

\begin{figure}[tbh]
  \centering
 \includegraphics[width=0.9\textwidth]{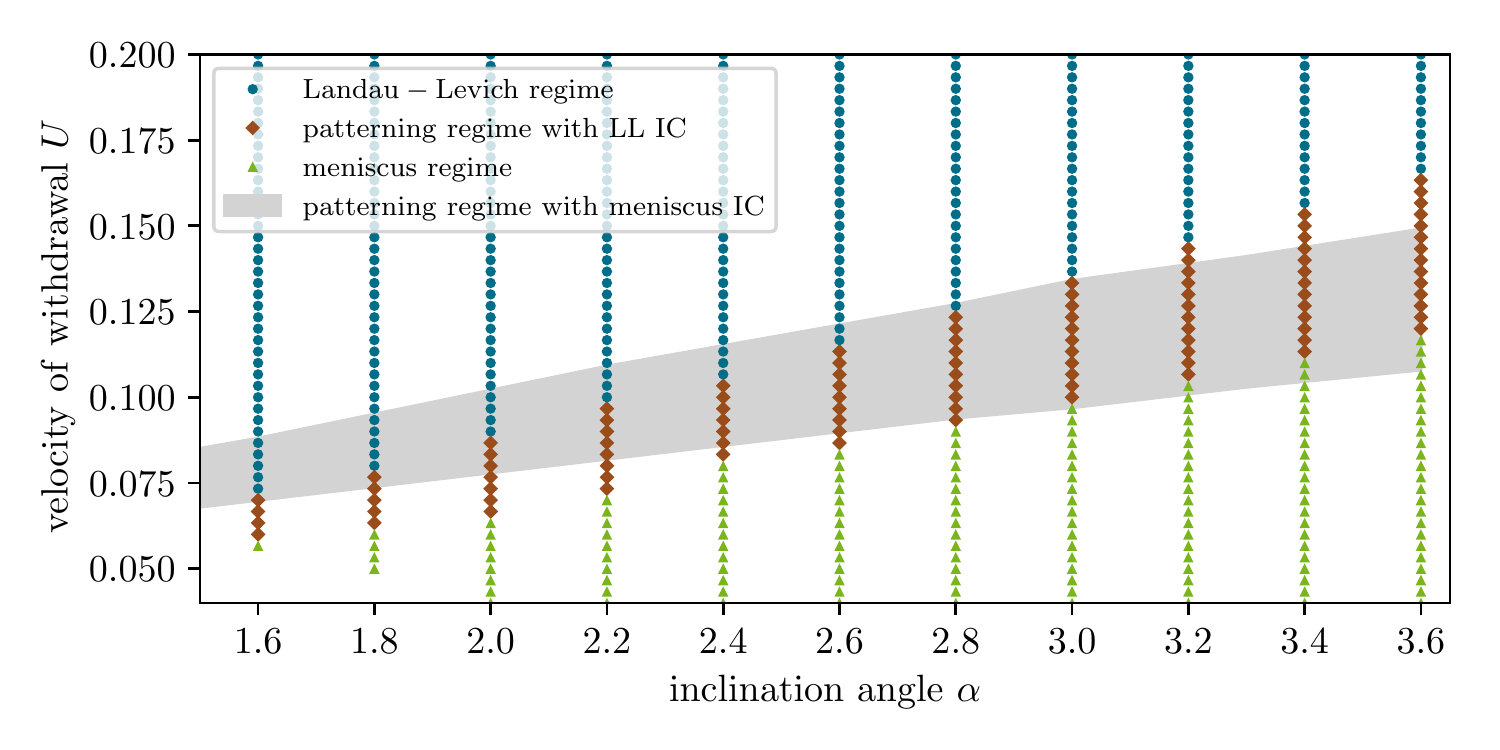}
 \caption{Morphological phase diagram showing where the different deposition regimes are found in the parameter plane spanned by the velocity of plate withdrawal $U$ and the inclination angle $\alpha$. The circle, square and triangle symbols indicate individual simulations starting from steady states obtained by numerical continuation. Namely, red squares stand for time-periodic states while blue filled circles and green filled triangles indicate steady Landau--Levich film deposition and steady meniscus states, respectively. The grey shaded area shows where periodic deposition is found if the simple profile \eqref{Eq:MeniscusIC} is employed as initial condition. The shaded area and the region of red squares do not perfectly match demonstrating the bistability of and hysteresis between steady and time-periodic states.}
 \label{fig:fig6_phase_diagram_1}
\end{figure}

 An overview of the behaviour in the $(\alpha,U)$ parameter plane is presented in the form of a morphological phase or state diagram in \reffig{fig:fig6_phase_diagram_1}. The red squares indicate the patterning region situated between regions of steady Landau--Levich film deposition (blue filled circles) and of steady meniscus states (green filled triangles). Each symbol represents a time simulation initialised by a steady solutions at the chosen pair $(\alpha,U)$ as obtained by continuation. However,  the borders of the patterning region depend on initial conditions indicating that the system exhibits bistability and hysteresis between steady and time-periodic states.
 For instance, the grey shaded area in \reffig{fig:fig6_phase_diagram_1} indicates the patterning regime obtained for a simple initial profile that interpolates between a straight free surfaces for both, the dragged film on the moving substrate and the liquid bath, i.e., the initial condition is
 \begin{equation}
  h(x) = \frac{1}{2}(h_l-h_\infty+\alpha x)\big(1-\tanh(x+(h_l-h_\infty)/\alpha) \big)+h_\infty.
  \label{Eq:MeniscusIC}
 \end{equation}
This initial condition is designed to match the boundary conditions at the left boundary, where $h(x=0)=h_l$ and $\left. \nabla h\right|_{x=0} = (\alpha,0)$ is demanded, as well as at the right boundary, where a flat film with height $h_\infty = 1$ is imposed.

Close inspection of the dynamics in the numerical time simulations leads us to the following interpretation of the mechanism of the observed pattern formation: Above but close to the Landau--Levich transition, linearly unstable flat films are transferred onto the moving plate. These linearly unstable films exhibit \textit{spinodal dewetting} if placed on a horizontal substrate \cite{Mitl1993jcis,ThVN2001prl,TVNB2001pre,LaCK2018jfm}. Here, the same surface instability is active if its timescale is comparable to the one determined by the velocity of withdrawal $U$ and the size of the meniscus region. Thus, the periodic deposition occurs as a result of the instantaneous dewetting of the deposited liquid layer near the meniscus region. 
Next, we more closely analyse the upper bound of the patterning region in the case where a well-defined steady Landau--Levich film is employed as initial condition of the time simulations.

\section{Marginal Stability Analysis}\label{SEC.MarginalStabilityAnalysis}

\begin{figure}[h!]
  \centering
 \includegraphics[width=0.9\textwidth]{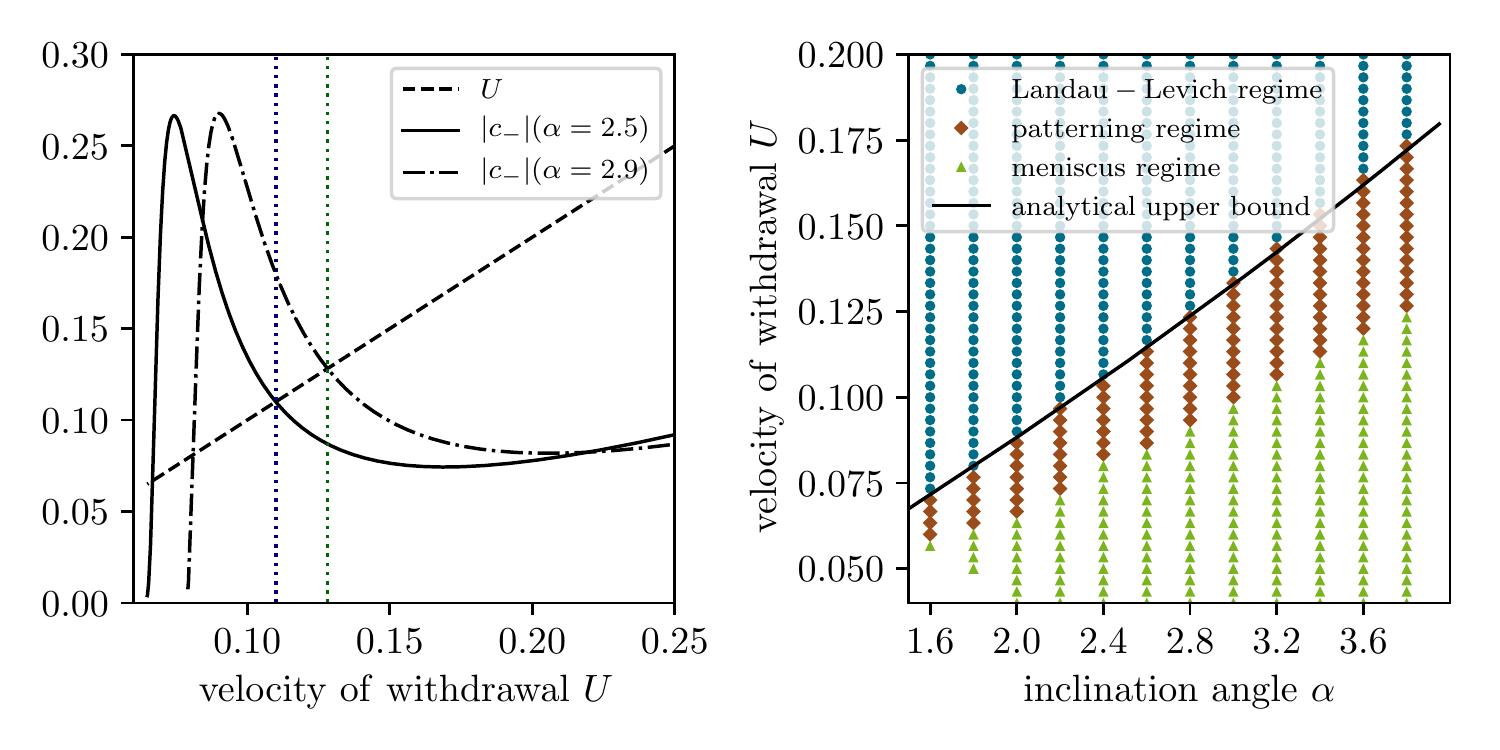}
 \caption{(left) Shown is for two different inclination angles the linear propagation speed $|c_{-}|$ of the surface instability that invades the unstable Landau--Levich film formed at given plate velocity $U$. The threshold velocities $U_{\mathrm{lim}}$, where $|c_{-}|$ equals $U$ are indicated by vertical dotted lines. Indeed, $U_{\mathrm{lim}}$ gives the \textit{upper} bound for pattern formation since for decreasing $U$, the speed $|c_{-}|$ increases. (right) State diagram showing the three deposition regimes as in \reffig{fig:fig6_phase_diagram_1} supplemented by a solid black line that indicates where $g(u)=|c_{-}(U)|-U$ crosses zero. It coincides well with the upper limit of the patterning region as obtained by numerical time simulations starting from solutions obtained by numerical continuation as initial conditions.}
 \label{fig:fig7_marginal_velocity_pahse_diagram_2}
\end{figure}

A well-established technique for the analytical calculation of the propagation speed of instability fronts that invade linearly unstable states is the \textit{marginal stability analysis} \cite{saarloos1988front,HuMo1990arfm,vanSaarloos2003}. In the context of thin-film flows similar analyses have been employed, e.g., in \cite{LJDT1992pd,LiKo2010pf,LaCK2018jfm}.

Here, we employ it to calculate the upper bound of the patterning region as it is given by the plate velocity at which the surface instability starts to invade the steady Landau--Levich film from the upstream end. First, we consider the propagation of a surface instability resulting in dewetting along flat unstable films $h_0(x)=h_0$ as described by the one-dimensional version of equation~\eqref{EQ.tfincline} omitting the advection ($U=0$). The basic assumption of the marginal stability analysis is that the propagation of an instability front into an unstable state is controlled by the dynamics linearised about this state. With other words, the instability front is \textit{pulled} by its leading edge that is well described by a linearization of the governing equations.

We introduce the constants 
\begin{align}
 \xi(h_0):=-f''(h_0)~~\mathrm{and}~~\eta(h_0):=\frac{\alpha G Q'(h_0)}{Q(h_0)}\label{EQ.marginalbasic}
\end{align}
and the rescaled dispersion relation \eqref{EQ.dispersionRelation}
\begin{align}
 \tilde{\omega}(k):=\frac{\omega(k)}{Q(h_0)}=\xi k^2-k^4+\eta ik. \label{EQ.marginalbasictwo}
\end{align}
Note that for $\alpha\neq0$ one has a non-zero imaginary part of the dispersion relation that indicates the occurrence of an oscillatory or travelling-wave instability. Here, it simply refers to a drift of the harmonic spinodal modes. The equations determining the scaled marginally stable linear propagation speed $\tilde{c}$ of a perturbation read \cite{vanSaarloos2003}:
\begin{align}
 \frac{\mathrm{Re}\left\{\tilde{\omega}(k^*)\right\}}{\mathrm{Im}\left\{k^*\right\}}=\tilde{c};~~~~  \mathrm{Re}\left\{\tilde{\omega}'(k^*)\right\}=0;~ -\mathrm{Im} \left\{\tilde{\omega}'(k^*)\right\}=\tilde{c}\label{EQ.marginalconst},
\end{align}
where the wavenumber $k^*$ is defined in the complex plane and the three unknowns are $\tilde{c},\ \mathrm{Im}\left\{k^*\right\}$ and $\mathrm{Re}\left\{k^*\right\}$. Since parity of the system ($x\to-x$ symmetry) is broken due to the inclination-related parameter $\eta$, we obtain two distinct marginal velocities $\tilde{c}_{+}$ and $\tilde{c}_{-}$. This is in contrast to systems where parity holds as, e.g., in the case of the decomposition of a binary mixture described by the Cahn--Hilliard equation \cite{Sche2017jdde}, for a film underneath a horizontal substrate destabilised by gravity described by a thin-film equation \cite{LJDT1992pd} or dewetting described by a thin-film equation \cite{LaCK2018jfm}.
The velocities are given by
\begin{align}
 \tilde{c}_{\pm}(h_0)=-\eta\pm\frac{2}{3}\xi^{3/2}\left[\left(2+\sqrt{7}\right)\sqrt{\frac{\sqrt{7}-1}{6}}~\right]. \label{EQ.marginalSpeed}
\end{align}
Here, $c_{-}=Q(h_0)\tilde{c}_-$ corresponds to a propagation towards the bath while $c_{+}=Q(h_0)\tilde{c}_+$ represents a propagation in the direction away from the bath. Returning to the full system \eqref{EQ.tfincline} at fixed inclination $\alpha$, where the plate is withdrawn with velocity $U>0$, we can now identify a threshold velocity $U_{\mathrm{lim}}$.  It is the velocity of withdrawal which exactly compensates for the speed $c_-$ of the instability moving towards the bath along the flat deposited film of height $h_\mathrm{f}$. However, here the height $h_\mathrm{f}$ is not an independent parameter but given as the thickness of the transferred Landau--Levich film at $U=U_{\mathrm{lim}}$.

In practice, to obtain the threshold velocity $U_{\mathrm{lim}}$ at fixed $\alpha$, we first calculate $h_\mathrm{f}$ as a function of $U$ by numerical continuation (see Section~\ref{SEC.LLTransition}). Then, we employ eq.~\eqref{EQ.marginalSpeed} to calculate $c_{-}(U)=c_{-}\left(h_\mathrm{f}(U)\right)$ and finally obtain the zero crossing of $g(U):=|c_{-}(U)|-U$ that gives $U_{\mathrm{lim}}$ as illustrated in \reffig{fig:fig7_marginal_velocity_pahse_diagram_2} (left).

\reffig{fig:fig7_marginal_velocity_pahse_diagram_2} (right) shows that these zero crossings at $U_{\mathrm{lim}}$ when calculated as a function of $\alpha$ (shown as a solid line) indeed very accurately predict the upper limiting velocity of the regime of periodic deposition. Note that an analogeous approach for the calculation of such limiting velocities has been employed in \cite{KGFT2012njp} for a reduced model system for pattern formation during Langmuir--Blodgett transfer. 
\begin{figure}[h!]
  \centering
 \includegraphics[width=0.9\textwidth]{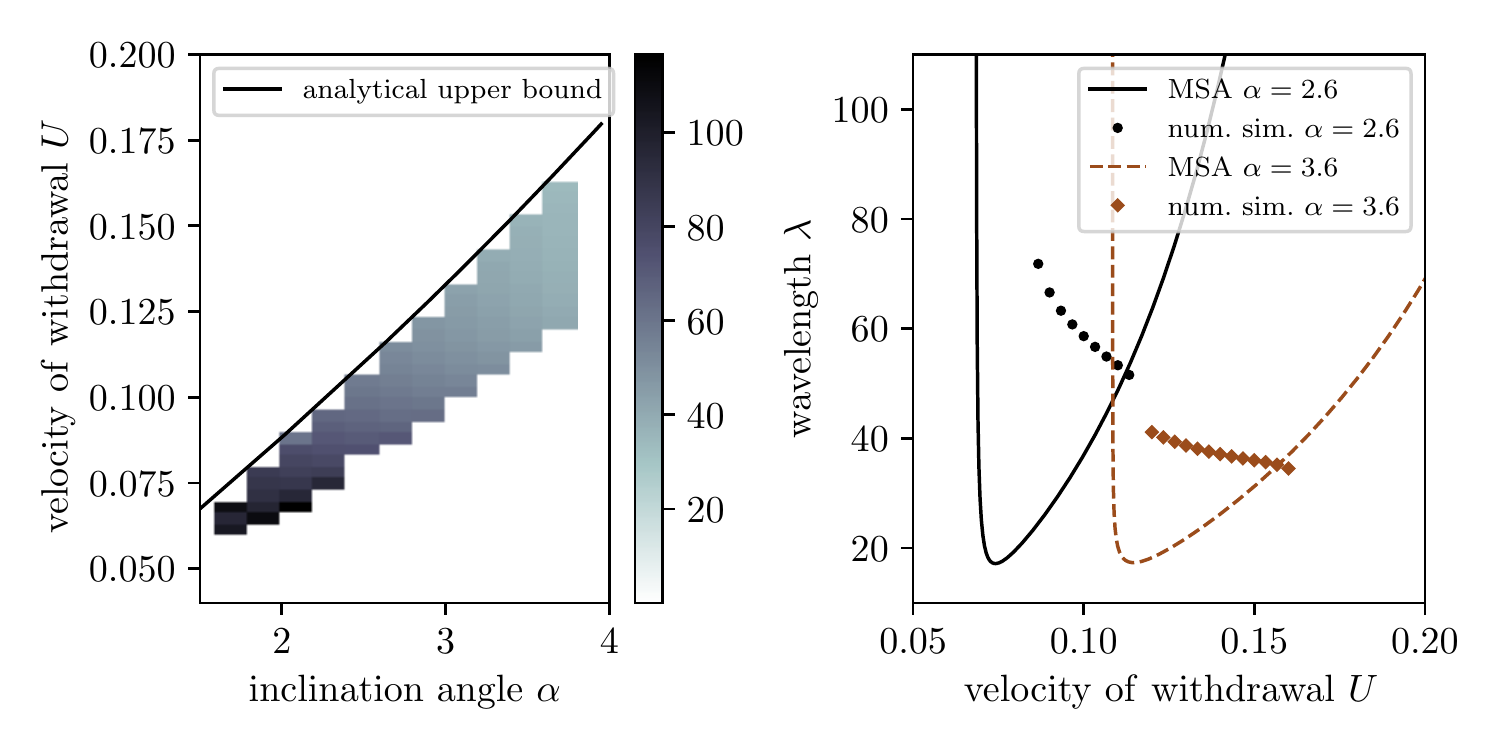}
 \caption{(left) Wavelength $\lambda$ of the deposited structures (coded in grey scale) in the $(\alpha, U)$-plane as extracted from numerical time simulations of equations \eqref{EQ.tfincline}-\eqref{EQ.bcupstream}.  The solid line indicates the instability threshold obtained from marginal stability analysis. (right) Wavelength of deposited structures from time simulations at two fixed inclinations $\alpha$ as given in the legend (filled square and circle symbols) and the analytically obtained wavelength selected by fronts marginally propagating into the corresponding flat Landau--Levich films (solid and dashed lines). As expected, the wavelengths coincide at the upper boundary of the patterning regime.}
 \label{fig:fig8_stripe_wavelength}
\end{figure}

In addition, one can calculate the wavenumber $k_s$ of the pattern behind the front which is selected by the leading edge of the patterning front \cite{saarloos1988front}, for the case without phase slips at the front. The calculation of the selected wavenumber $k_s$ behind the front is based on the assumption that the flux of nodes present in the oscillatory leading edge of the instability front with the wavenumber $k^*$ obtained from \eqref{EQ.marginalbasic} is conserved across the front \cite{DeLa1983prl}. From this consideration, one obtains
\begin{align}
 k_s={\mathrm{Re}\left\{k^*\right\}}+\frac{\mathrm{Im}\left\{\tilde{\omega}(k^*)\right\}}{c}.
\end{align}

Here, in the dip-coating geometry, the instability front moves into the system from the domain boundary on the upstram side of the system, i.e., where the Landau--Levich film is dragged out of the domain.
The analytically determined wavenumber can only be expected to describe the periodic structures formed close to the onset of the instability well away from the meniscus region. This is only the case at the upper limit of the patterning regime, which is	 well described by the marginal stability analysis. 
We have extracted the emerging wavelength $\lambda$ from our numerical time simulations and present them in \reffig{fig:fig8_stripe_wavelength} (left) in the $(\alpha, U)$-plane where the black solid line again indicates the analytically obtained instability threshold. The right panel of \reffig{fig:fig8_stripe_wavelength} compares the wavelength $\lambda$ obtained from  for two fixed values of $\alpha$ as a function of $U$ with the full curve of results $\lambda_s(U)=2\pi/k_s(U)$ of the marginal stability analysis. Indeed, the two wavelengths well agree at onset, i.e., on the boundary of the patterning region, but deviate at smaller $U<U_{\mathrm{lim}}$. Note, that the curve corresponding to time simuations for $\alpha=2.6$ hints at a strong increase in $\lambda$ for smaller $U$. This could indicate a divergence of $\lambda$ as seen at a global bifurcation. To further discuss this hypothesis and to better understand the hysteresis between steady and time-periodic states, we next perform a numerical bifurcation analysis that includes time-periodic states.



\section{Bifurcation Structure of Periodic Solutions}
\label{SEC:BifStuctureOfPeriodic}
Up to here, we have beside numerical time simulations and analytical linear considerations employed numerical pseudo-arclength continuation to determine bifurcation diagrams of steady states as presented, e.g., in Fig.~\ref{fig:fig3_bifurcation_LL}. To do so, we have reformulated the 1D steady version of \eqref{EQ.tfincline} as a system of three first order ODEs in space. This has allowed us to use the auto07p-mode for a boundary value problem (BVP) that comes with a fully adaptive spatial discretisation and, therefore, facilitates consideration of relatively large domains.
However, to obtain a more complete picture of the behaviour close to the Landau--Levich transition, in particular, close to the onset of periodic deposition where 
bistability and hysteresis can occur (cf.~\reffig{fig:fig6_phase_diagram_1}), we now pursue the numerical continuation of time-periodic solutions. As the toolbox auto07p supports continuation of time-periodic solutions only for systems of ODEs, we discretise Eq.~\eqref{EQ.tfincline} ``by hand'' with a finite difference discretisation employing a fixed equidistant grid of $N$ nodes in space resulting in $N$ coupled ODEs in time. The full auto07p-intrinsic adaptive grid is then employed for the discretisation in time for the time-periodic solutions only. This approach has also been employed in \cite{KoTh2014n} for the numerical continuation of periodic solutions of a Cahn--Hilliard model for Langmuir--Blodgett transfer experiments. For the continuation of periodic solutions of a similar problem with periodic boundary conditions, a spatial discretisation by pseudo-spectral methods is performed in \cite{LRTT2016pf}. For a recent review see \cite{EGUW2018preprint}.
Since the continuation of periodic solutions of $N$ ODEs is for large $N$ computationally very expensive, we adapt our  system parameters accordingly to $h_l=40,~L=200$, i.e., move to a smaller system than the one studied before. The system is still large enough that we expect to find all features observed before. We choose a grid with $N=300$ points and tested that a larger value for $N$ does not change the location of the bifurcations. The steady solutions obtained with the BVP approach are reproduced in a good approximation.

\begin{figure}[h!]
  \centering
\includegraphics[width=0.75\textwidth]{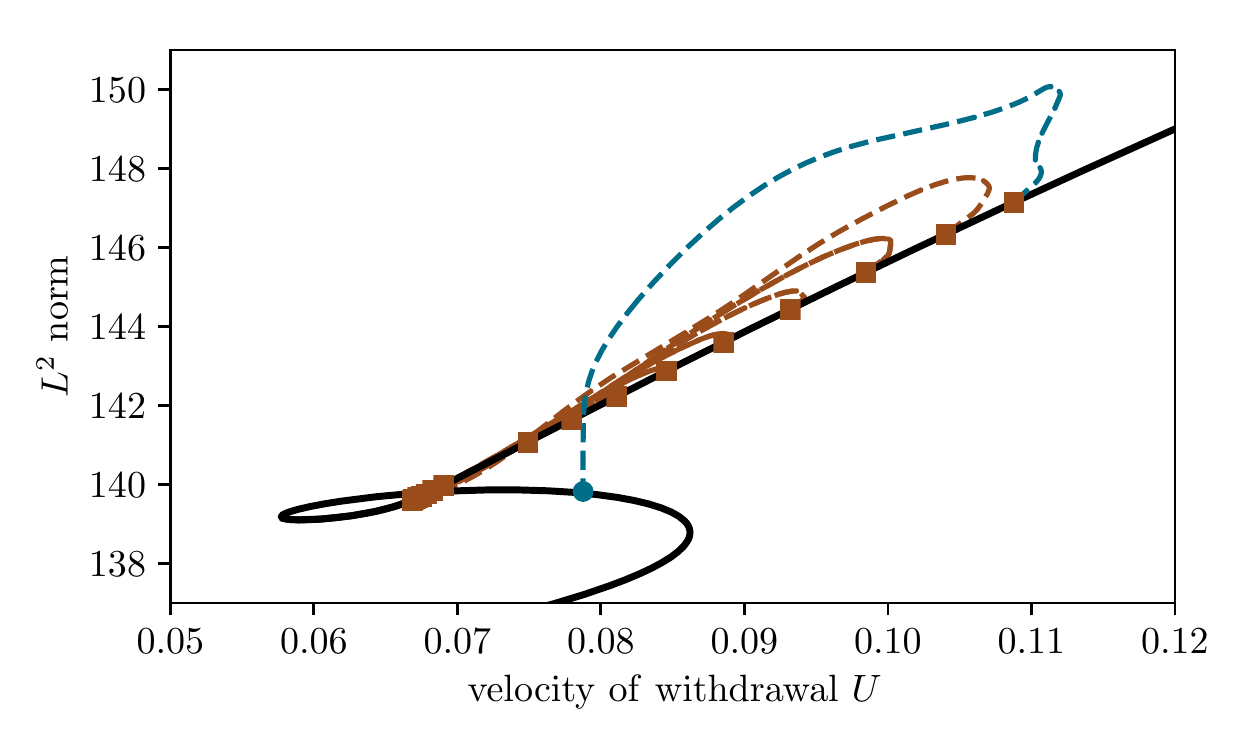}
\caption{Bifurcation diagram for $\alpha=2.0$ close to the Landau--Levich transition. The branch of steady states is shown as solid black line and branches of time periodic states are given as dashed lines. Most of the latter emerge via subcritical Hopf bifurcations (filled red squares). All red dashed branches connect two Hopf bifurcations. Only the branch emerging at the Hopf bifurcation at the largest velocity $U$ terminates in a global bifurcation (filled blue circle). The solution measure is the $L^2-$norm of the solutions. For time-periodic states, it is averaged over one period. Parameters are $\alpha=2.0$ $h_l=40,~L=200$ and $N=300$.}
 \label{fig:figY_Bif_Periodic_Full}
\end{figure}

\begin{figure}[h!]
  \centering \includegraphics[width=0.9\textwidth]{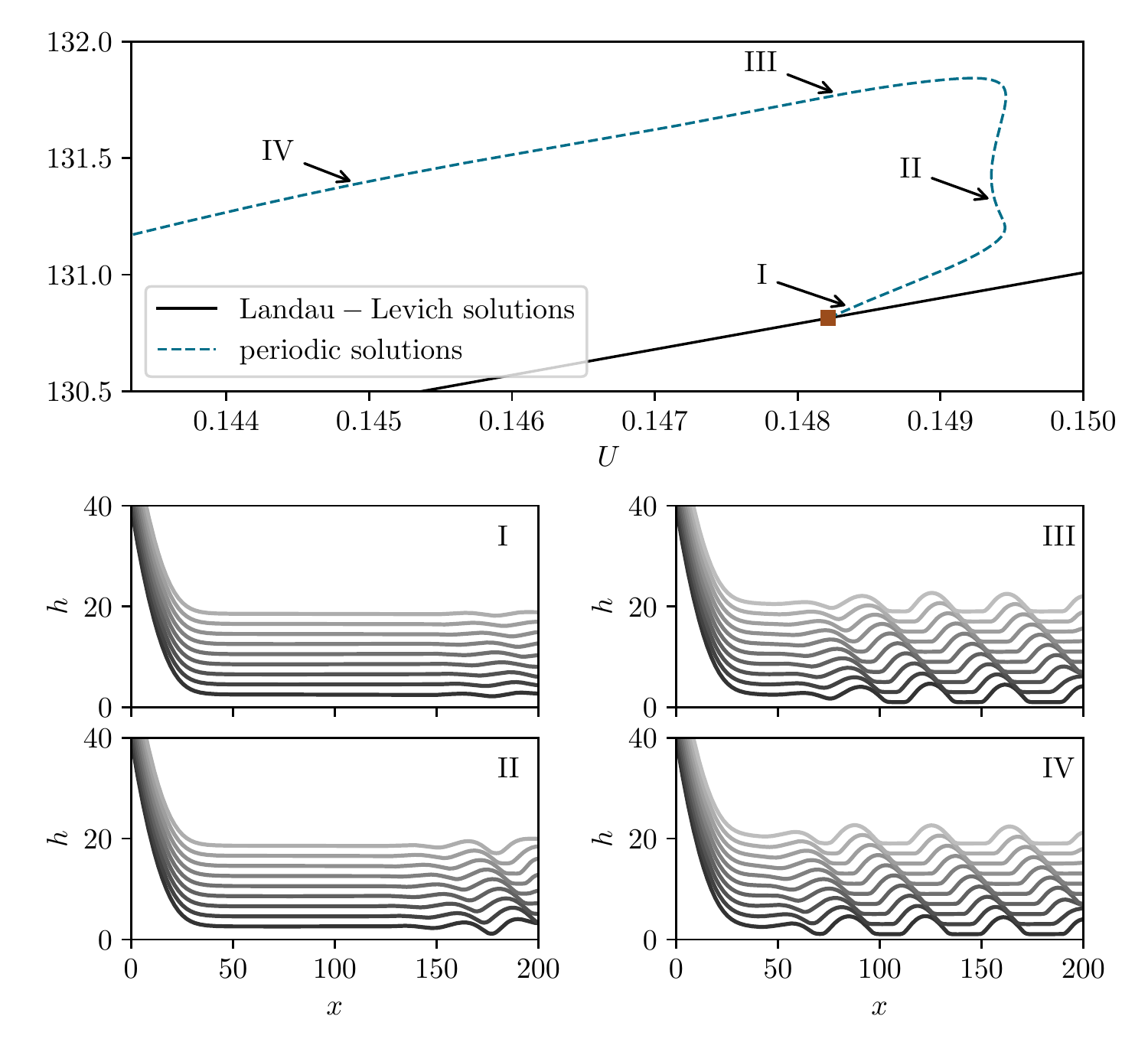} \caption{(top) Bifurcation diagram showing the branch of steady Landau--Levich films (solid line) and the dominant branch of time-periodic states (blue dashed line) in the region where hysteresis occurs. The Hopf bifurcation where the latter emerges is marked by a red filled square symbol. (bottom) Space-time plots consisting of staggered snapshots are shown in the four lower panels at loci indicated by roman numbers I to IV along the branch of time-periodic states. The time evolution is visualised by a vertical offset of  $h\approx 2$ between subsequent snapshots that lie $\Delta t\approx 10$ apart. Parameters are $\alpha=2.5$, $h_l=40,~L=200$ and $N=300$.}
 \label{fig:figX_Bif_Periodic_snapshots}
\end{figure}

The resulting bifurcation diagram for $\alpha=2.0$ is shown in \reffig{fig:figY_Bif_Periodic_Full}. It well illustrates the intricate pattern formed by the branch of steady states and a number of branches of time-periodic states. Starting at large velocity $U$, we decrease $U$ and follow the branch of steady Landau--Levich films. Before reaching the saddle-node bifurcation at $U\approx0.058$, we find a series of 17 Hopf bifurcations. The 17 bifurcations form the starting points of 9 branches of time-periodic states. Each of the 8 branches (red dashed lines in \reffig{fig:figY_Bif_Periodic_Full}) connects two of the Hopf bifurcations and consists of unstable time-periodic states.
In contrast, the branch emerging at the first Hopf bifurcation (at $U\approx0.11$, blue dashed line in \reffig{fig:figX_Bif_Periodic_snapshots}) connects nearly vertically with the unstable branch of steady states connecting the two saddle-node bifurcations that corresponds to the detached foot solutions discussed in Sec.~\ref{SEC.LLTransition} and shown in \reffig{fig:fig4_bifurcation_LL_foot}. The time period diverges upon approach of the steady state indicating that this branch of time-periodic states ends in a homoclinic global bifurcation \cite{Strogatz2014}.
A partly similar general picture is also found for the aforementioned Cahn--Hilliard model of Langmuir--Blodgett transfer \cite{KoTh2014n}, where the authors termed it a \textit{harp-like} bifurcation structure. 

The branch of time-periodic states bifurcating at the highest value of $U$ emerges subcritically towards the stable Landau--Levich films at higher $U$, i.e., it consists of unstable states. It then passes a first saddle-node bifurcation where it turns back towards smaller $U$ and becomes linearly stable. A second saddle-node bifurcation at slightly lower $U$ destabilises the branch again that now continues again towards larger $U$.  This brief interlude becomes slightly more pronounced at larger $\alpha$ (see \reffig{fig:figX_Bif_Periodic_snapshots} (top)). At a third saddle-node bifurcation, the branch again turns back towards smaller $U$ and becomes linearly stable. It remains linearly stable until ending in the global bifurcation. The stable part of the branch corresponds to the line deposition commonly found in the discussed time simulations. The extended subcritical region between the primary Hopf bifurcation at $U=0.1088$ and the third saddle-node at $U=0.1121$ explains the bistability and hysteresis found in the time simulations.

\begin{figure}[h!]
  \centering
\includegraphics[width=0.75\textwidth]{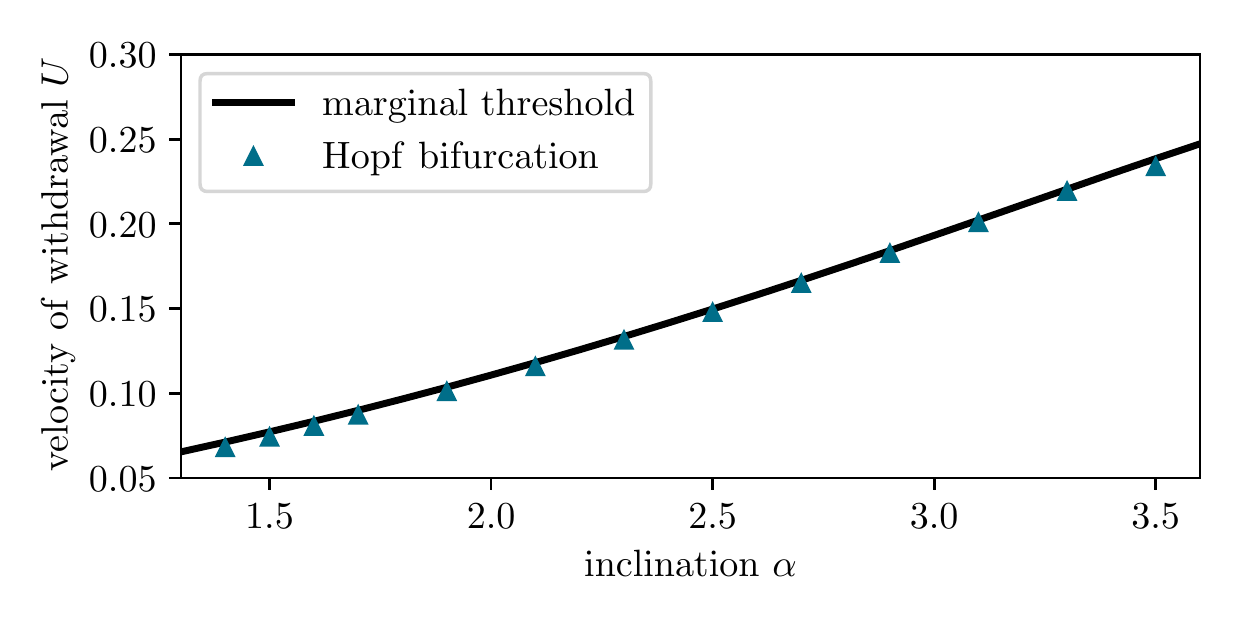}
\caption{Comparison of the loci of the Hopf bifurcation where the dominant branch of time-periodic states emerges and of the instability threshold obtained via the marginal stability analysis described in Sec.~\ref{SEC.MarginalStabilityAnalysis} in the plane spanned by inclination $\alpha$ and plate velocity $U$.
}
 \label{fig:figZ_hopf_vs_marginal}
\end{figure}

In \reffig{fig:figX_Bif_Periodic_snapshots}, we show focus on the region where hysteresis occurs between steady and time-periodic states and show the bifurcation diagram for $\alpha=2.5$ together with examples of space-time plots of unstable and stable time-periodic solutions. Close to the primary Hopf bifurcation, the time-periodic states correspond to small-amplitude oscillations close to the right hand end of the domain (set I). By eye, only two spatial periods can be distinguished. This behaviour is consistent with the interpretation that at the Hopf bifurcation point, the leading edge of an instability starts to propagate from the right hand boundary towards the meniscus counteracted by the advection term. With other words, the Hopf bifurcation obtained through numerical continuation should nearly coincide with the stability border obtained via the marginal criterion outlined in Sec.~\ref{SEC.MarginalStabilityAnalysis} evaluated for the film height of the deposited Landau--Levich film. \reffig{fig:figZ_hopf_vs_marginal} shows that this is indeed a very good approximation across a range of inclination angles $\alpha$.

However, the time-periodic solutions on the first subbranch, i.e., the first subcritical part of the branch emerging at the Hopf bifurcation, are linearly unstable. This might be explained by considering a marginal propagating front into an inhomogeneous steady state (corresponding to the film height away from the meniscus that decays towards the Landau--Levich film height). Similar theoretical concepts are used in hydrodynamics in the context of so called global modes that occur, in particular, in open flow systems with broken Galilean invariance \cite{CoCh1997pd,CoCh1999pd}, a property shared by the present dragged film system. In time simulations initialised with individual snapshots from an unstable time-periodic state such as state I in \reffig{fig:figX_Bif_Periodic_snapshots}, the instability front typically propagates towards the meniscus and the system subsequently relaxes to a time-periodic state corresponding to the uppermost subbranch shown in \reffig{fig:figX_Bif_Periodic_snapshots}, i.e., where states III and IV are located. 

 Following the branch of time-periodic states from state I through the three saddle-node bifurcations and then towards lower velocities of withdrawal (passing states II, III, and IV of \reffig{fig:figX_Bif_Periodic_snapshots}, the height modulations become stronger transforming into a periodic array of ridges. In parallel, the spatial onset of the modulations moves continuously closer to the meniscus. Finally, at sufficiently small velocities $U$, the ridges are formed directly at the meniscus (see state IV in \reffig{fig:figX_Bif_Periodic_snapshots}). In this limit one can say that an oscillating meniscus emits ridges that are transported away by the moving plate. This is another reason why a marginal stability analysis of a homogeneous state is not able to give the lower instability onset.

\section{Two-Dimensional Flows}\label{SEC.2D}
\begin{figure}[h!]
  \centering
 \includegraphics[width=0.75\textwidth]{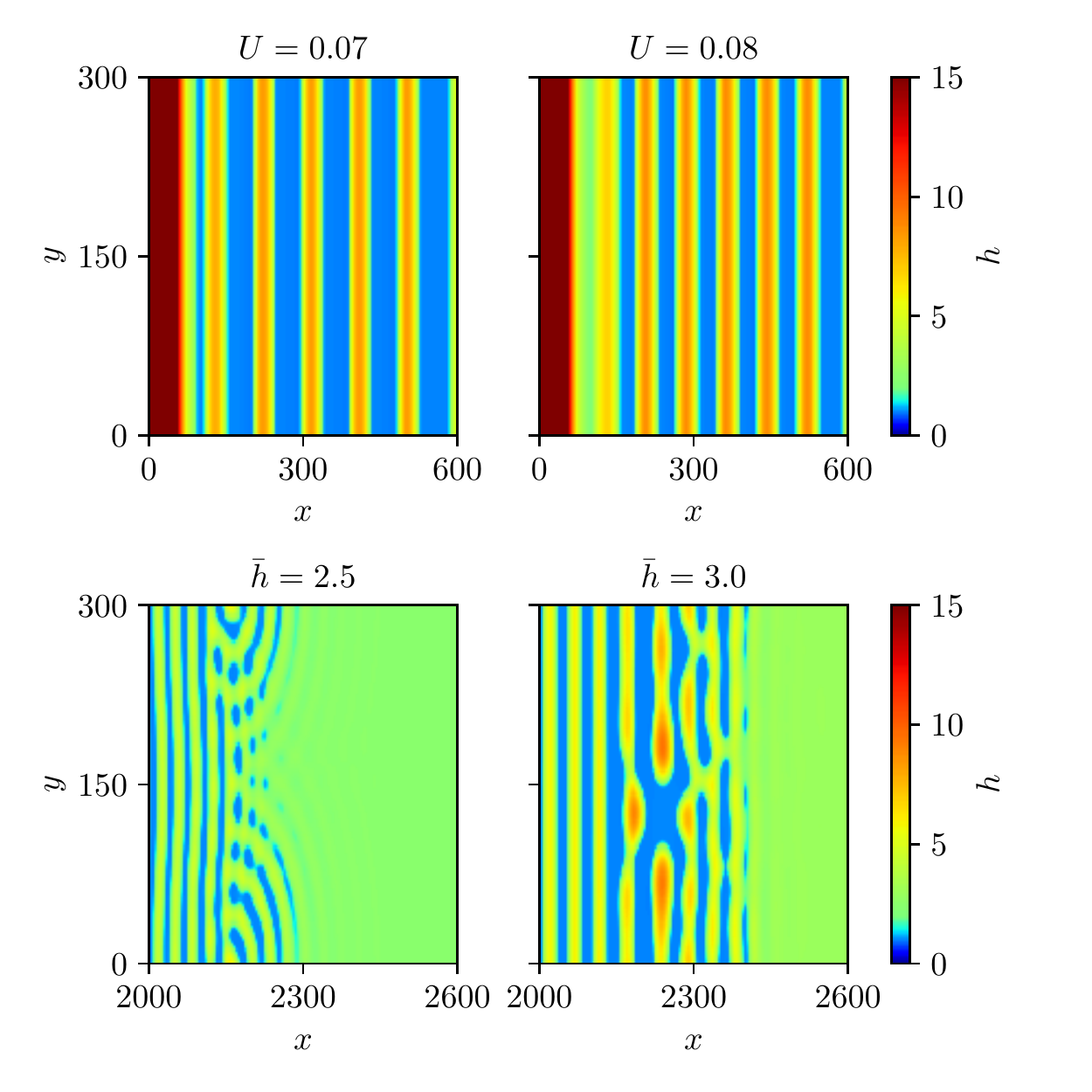}
 \caption{(top) Shown are snapshots from numerical simulations of stripe formation in the 2D dip-coating geometry after initial transients have died out for two velocities (left) $U=0.07$ and (right) $U=0.08$ at inclination $\alpha=2.0$. The domain size is $L_x\times L_y=300\times 600$ and the film height at the boundary on the meniscus side (l.h.s. of panel) is $h_l=100$. About 150 stripes have already left the domains at their r.h.s.. No secondary instability is visually discernible. (bottom) Shown are snapshots from the directed dewetting of flat films for two different mean film heights (left) $\bar{h}=2.5$ and (right) $\bar{h}=3.0$ (approximately corresponding to $h_\mathrm{f}$ at the velocities in the top images) initialised with a straight front at $x=0$. The computational domain size is $L_x\times L_y =300\times 3000$. In $x$-direction only part of the simulation domain is shown, i.e., most of the initially formed stripes are not shown ($\sim 50$).
The dewetting fronts that initially form stripes clearly undergo a transversal instability.}
 \label{fig:fig9_snapshots_2D}
\end{figure}

Up to here, all our considerations have concerned 1D film flows, i.e., we have assumed that all structures are translationally invariant in the direction orthogonal to the direction of dragging ($x$-coordinate). In the final section we consider 2D dragged films. 

It is known that on horizontal substrates in the absence of lateral driving terms, fronts of spinodal dewetting are often transversally unstable \cite{ReSh2001prl,NJSB2003jpcm,BeTh2010sjads}.
Our numerical time simulations suggest that this is not the case in the considered dip-coating geometry as the dragging of the plate stabilises the transversal instability modes. In \reffig{fig:fig9_snapshots_2D}, we show snapshots of time simulations for (bottom panels) 2D substrates for a dewetting front on a horizontal substrate and (top panels) for a dragged film.
For the dip-coating geometry, i.e., for non-zero inclination angles and dragging velocities, the deposition of regular straight stripes still persists after a long time (\reffig{fig:fig9_snapshots_2D} (top)) indicating that transversal ridge instabilities \cite{ThKn2003pf} upon formation of the stripes are either completely stabilised or so weak that we do not observe them in the finite computation time.

In contrast, a standard dewetting front on a horizontal substrate is clearly transversally unstable, see \reffig{fig:fig9_snapshots_2D} (bottom) where such a front is initiated at the left hand end of the domain and moves towards the right. Note that both simulations in the bottom row start with an initially flat film of a height corresponding approximately to the height $h_\mathrm{f}$ obtained for the respective set of parameters in the dip-coating geometry, i.e., the two dragging velocities in the top row of \reffig{fig:fig9_snapshots_2D} result in two specific $h_\mathrm{f}$ that are also employed in the bottom row.
For both initial film heights, first the front is nearly transversally invariant but soon develops a strong transversal instability. This occurs even without the addition of explicit noise to the initial conditions. Even without conducting a transient transversal stability analysis of stripe-forming fronts, the time simulations clearly suggest that the stripe deposition process in dip-coating is significantly less unstable due to the presence of strongly symmetry-breaking directional forcing. Note, however, that each deposited stripe is, in general, unstable with respect to a Plateau--Rayleigh-type instability (cf. \cite{thiele2003modelling,diez2009breakup}) that acts, however, on a larger time scale.

\section{Conclusion and outlook}
\label{sec:conc}
In the present work, we have employed a hydrodynamic thin-film model to numerically and analytically investigate the withdrawal of a solid substrate from a bath of nonvolatile partially wetting simple liquid. We have predicted that in this classical dip-coating geometry, there exists a third withdrawal regime beside the two well-known ones \cite{TeDS1988rpap} where the substrate is either (i) macroscopically dry (covered by a microscopic adsorption layer) or (ii) homogeneously coated by a macroscopic liquid Landau--Levich film whose thickness depends on the angle and velocity of substrate withdrawal. The third regime consists in (iii) the deposition of a regular pattern of liquid ridges that are oriented parallel to the meniscus. Although the possibility of an instability of the deposited Landau--Levich film is mentioned in the conclusion of Ref.~\cite{TeDS1988rpap}, to our knowledge the regime was not analysed before as part of the solution and bifurcation structure of the long-wave description of dip-coating processes. 

In contrast to thin-film slip models of dip-coating \cite{SZAF2008prl,ZiSE2009epjt} that incorporate partial wettability directly through an imposed finite contact angle at the contact line (in slip models the point where the film height strictly goes to zero), here we have employed a precursor-film model \cite{GTLT2014prl}. It includes a Derjaguin (or disjoining) pressure that guarantees that even a macroscopically dry region is covered by a microscopic adsorption layer and also accounts for the correct macroscopic equilibrium contact angle. Employing such a model has allowed us to investigate the interplay of the dynamic deposition of the film and a surface instability of the deposited Landau--Levich film that is related to spinodal dewetting \cite{Mitl1993jcis,ThVN2001prl,LaCK2018jfm}. This interplay provides the mechanism of the meniscus instability that results in the periodic shedding of ridges that form a regular deposition pattern. One may say that the pattern results due to instantaneous dewetting of the coating film as it emerges from the meniscus. 

Our detailed analysis has combined a marginal stability analysis of a dewetting front advancing into an unstable flat film, time simulations in the fully nonlinear regime and a bifurcation study combining two types of numerical path-continuation. Thereby, we have shown that the marginal stability analysis determines accurately the upper limiting velocity observed in numerical time simulations.
This analysis determines the threshold substrate velocity below which a linear dewetting instability front that moves from the upstream boundary of the domain into the deposited Landau--Levich film (i.e., towards the meniscus) is not anymore advected away by the moving plate. Furthermore, the fully nonlinear simulations have shown that regular patterns of deposited ridges can also be found in a certain region \textit{above} this upper threshold velocity. 
This implies that there is a range of velocities where the system shows bistability between homogeneous and periodic deposition, i.e., hysteresis effects.

The detailed numerical bifurcation analysis revealed a number of Hopf bifurcations on the solution branch describing a homogeneous transfer of which the one occurring at highest velocity corresponds to the upper limiting velocity determined by the marginal stability analysis. Together they result in the emergence of an intricate structure formed by branches of time-periodic states. This solution structure exhibiting sub-critical Hopf bifurcations well explains the existence of bistability and hysteresis observed in the numerical time simulations. Note that somewhat similar structures are found in other related systems where a spatial heterogeneity (here, the meniscus where the bath ends) interacts with a lateral driving force (here, the withdrawal of the plate) in the vicinity of a phase transition (here, the wetting transition occurring when the equilibrium contact angle goes to zero and the dewetting instability ceases to exist). For instance, a more complicated bifurcation structure is found in a study of Langmuir--Blodgett transfer of a layer of surfactant employing a model based on a convective heterogeneous Cahn--Hilliard equation \cite{kopf2012substrate,KoTh2014n}. There, many branches of time-periodic concentration profiles emerge from an intricately snaking branch of steady profiles. The branches of time-periodic states represent the periodic transfer of stripes of different surfactant phases and normally connect a Hopf bifurcation and a global (sniper or homoclinic) bifurcation. Overall, they form a harp-like structure that is topologically similar to our Fig.~\ref{fig:figY_Bif_Periodic_Full}. Note, that in both cases the number of Hopf bifurcation points will depend on system size as more modes fit into larger domains.

Note, that the understanding of the bifurcation scenario of deposition processes is not only important for the dip-coating process at hand and the Langmuir--Blodgett transfer \cite{kopf2012substrate,KoTh2014n}. It is also relevant for line deposition from solutions and suspensions with volatile solvent where other instability mechanisms dominate, see, e.g., the case of evaporative dewetting \cite{FrAT2012sm,DoGu2013el}. It was pointed out in \cite{FrAT2012sm} and more extensively discussed in \cite{KoTh2014n} that the onset of periodic deposition at the lower limiting velocity may be considered a depinning transition as beyond this critical velocity part of the steady meniscus profile depins and is dragged away from the bath as liquid ridge.
This relates our result to other depinning transitions, e.g, of driven drops on substrates with defects \cite{BKHT2011pre,VFFP2013prl}, of drops on rotating cylinders \cite{LRTT2016pf}, and of interface undulations of an air finger in a liquid-filled channel from its advancing tip \cite{ThJH2014jfm}. In most cases, depinning is triggered at global (sniper or homoclinic) bifurcations and the emerging branches of time-periodic states normally end in Hopf bifurcations at large driving velocities. Also note that the combination of an imposed plate withdrawal velocity and a heterogeneity (meniscus fixed in the lab system) in our dip-coating system shows similarities with a quenching front moving with an imposed velocity. The latter may produce a variety of patterns (for a Cahn--Hilliard-type system, see Ref.~\cite{FoWa2009pre,FoWa2012pre}) as mathematically investigated in detail in Ref.~\cite{GoSc2015arma}.

Finally, we focus on implications of our results for dip-coating, other coating processes and related hydrodynamic systems. The underlying mechanism of the ridge deposition is universal and will apply to all dragged-film and other coating systems when the deposited coating film is linearly unstable, independently of the instability mechanism and geometry. This includes, e.g., films of dielectric liquids destabilised by an electrical field (see appendix of Ref.~\cite{ThKn2006njp}), liquid films deposited on a heated substrate that are unstable w.r.t.\ a long-wave Marangoni instability (as discussed in Refs.~\cite{OrDB1997rmp,ThKn2004pd}), partially wetting liquids driven by surface accoustic waves (systems as in Refs.~\cite{MZAM2016l,MoMa2017jfm} but using partially wetting liquids) and coating processes for a fiber \cite{Qu1999arfm}.

Further note that our results are also relevant for the original Bretherton problem of a gas bubble that moves through a liquid filled tube \cite{Bret1961jfm,TeDS1988rpap}. There, recent experiments \cite{KSPK2018prf} investigate long bubbles that move in rectangular channels filled with partially wetting liquid (also cf.~\cite{DrTW2003prl,JoCu2014ra}). They describe three distinct regimes: (i) a fully dewetted regime without a visible film, (ii) a fully wetted regime where the deposited film is stable on the convective timescale of the bubble and (iii) an intermediate regime where the deposited film ruptures and a sequence of droplets emerges. The three regimes closely correspond to the three regimes described here, i.e., our study also provides an outline of the basic bifurcation structure for the Bretherton problem for partially wetting liquids. A somewhat similar transition related to the transition between our regimes (ii) and (iii) was recently described in connection with the dynamics of relatively thick viscous liquid films flowing down a cylindrical fiber \cite{JFSZ2019jfm}. 
Of interest in the context of the present study is their transition between a Rayleigh--Plateau regime, where a regular sequence of relatively small drops connected by short stable liquid films slides down the fiber (stable travelling wave), and an isolated droplet regime where widely spaced larger drops slide down the fiber connected by an unstable coating film that develops into a number of small slower droplets. In the frame moving with the large drops small droplets are shed from their trailing meniscus. The transition occurs with increasing drop size and distance as larger drops slide faster and leave a thicker coating layer behind. This layer becomes linearly unstable above a critical thickness and the small droplets develop in analogy to the time-periodic shedding of ridges in our case. This implies that the here developed methodology based on the combination of marginal stability analysis and numerical bifurcation study could prove fruitful for the mentioned systems.

The main part of our study has focused on basic mechanisms and the bifurcation structure for one-dimensional substrates, i.e., for an imposed translational invariance in the direction orthogonal to the direction of withdrawal.  However, we have also employed numerical time simulations to test the validity of our main results in the full two-dimensional geometry. We have indeed found that the symmetry-breaking forcing introduced by the plate withdrawal in the dip-coating geometry significantly stabilises the formation of stripes as compared to the case of an initially straight dewetting front that recedes without additional forcing on an isotropic homogenous substrate.
In the future, it would be desirable to deeper investigate the two-dimensional geometry, i.e., to perform transient transversal linear stability analyses of the one-dimensional solutions as well as numerical bifurcation analyses via continuation techniques. The linear analysis should establish whether there exists a parameter region where the formed ridge patterns are indeed stable with respect to transversal perturbations on any time scale. An investigation of the bifurcation structure should be embedded into a more general analysis of the bifurcation structures of spatially one- and two-dimensional time-periodic states for several of the above described systems resulting in pattern deposition and depinning.

\begin{acknowledgments}
	SG and WT acknowledge support by Deutsche Forschungsgemeinschaft (DFG) within PAK 943 (Project No. 332704749). UT thanks the DFG for funding under Grant No.~TH781/8-1.
\end{acknowledgments}

\bibliography{literature}

\end{document}